\documentclass[12pt,a4paper]{elsarticle}
\usepackage{ifpdf}
\usepackage[%
  \ifpdf\else hypertex,\fi%  To keep YAP happy, remove from DVIPS
  backref,citecolor=blue,%
  pdftitle={Numerical determination of partial spectrum of Hermitian matrices
    using a Lanczos method with selective reorthogonalization},%
  pdfauthor={Chris Johnson and A. D. Kennedy},%
  pdfkeywords={Spectrum, Lanczos, Krylov, Eigenvalue, Eigenvector, Hermitian,
    LANSO, Lattice},% 
  pdfstartview={FitH},%
  pdfpagemode={FullScreen}]{hyperref}
\usepackage{bm}
\usepackage{bbm}
\usepackage{epsfig}
\usepackage{euscript}
\usepackage{caption}
\def\bb{\mathbbm}
\newif\ifdraft \draftfalse
\def\note[#1]#2{\message{(#1)}\ifdraft Y{\noindent\em[#2]\/}\fi}
\ifdraft
\fi
\def\epspdffile#1{\leavevmode\ifpdf\epsffile{#1.pdf}\else\epsffile{#1.eps}\fi}
\newcommand{\Lanczos}{L\'anczos}
\newcommand{\lanczos}{L\'anczos}

\newcommand{\ie}{{\em i.e.,\ }}
\newcommand{\mtx}[1]{{\mathsf #1}}

\def\Span(#1){\mathop{\rm span}\nolimits(#1)}
\def\Krylov(#1,#2,#3){{\mathcal K}_{#3}(#1,#2)} % Krylov space
\def\binomial(#1,#2){\left(\begin{array}{c} #1 \\ #2 \end{array}\right)}
\def\O{{\cal O}}		      % Big-O
\def\ulp{\varepsilon}		      % Unit of least precision
\def\binomial(#1,#2){\left(\begin{array}{c} #1 \\ #2 \end{array}\right)}
\def\defn{\equiv}                   % definition.
\newcommand{\ma}[1]{{\mtx{#1}}}
\newcommand{\QR}{QR}
\def\R{{\bb R}}			      % Set of real numbers
\def\eqref#1{(\ref{#1})}            % Equation cross reference
\def\secref#1{\S\ref{#1}}           % Section cross reference
\def\figref#1{Figure~\ref{#1}}      % Figure cross reference
\newcommand{\vect}[1]{{\mathbf #1}}
\newcommand{\KAPPA}{\(\kappa\)}	% KAPPA bound procedure
\def\emphname#1{\emph{#1}}
\def\emphdef#1{\emph{#1}}
\begin{document}
%
% Title page
%
\title{Numerical determination of partial spectrum of Hermitian matrices using
  a \Lanczos\ method with selective reorthogonalization}
\author{Chris Johnson\footnote{chrisj@epcc.ed.ac.uk}
  \hskip1.3em and\hskip1.3em
  A.~D.~Kennedy\footnote{adk@ph.ed.ac.uk} \\[1ex]
  SUPA, NAIS, and EPCC, \\
  Department of Physics and Astronomy, \\
  The University of Edinburgh, The King's Buildings, \\
  Edinburgh, EH9~3JZ, Scotland}
\begin{abstract}
  \noindent We introduce a new algorithm for finding the eigenvalues and
  eigenvectors of Hermitian matrices within a specified region, based upon the
  LANSO algorithm of Parlett and Scott.  It uses selective reorthogonalization
  to avoid the duplication of eigenpairs in finite-precision arithmetic, but
  uses a new bound to decide when such reorthogonalization is required, and
  only reorthogonalizes with respect to eigenpairs within the region of
  interest.  We investigate its performance for the Hermitian Wilson--Dirac
  operator \(\gamma_5D\) in lattice quantum chromodynamics, and compare it
  with previous methods.{\parfillskip=0pt\par}\vskip2ex
  \noindent\emph{Keywords:} Spectrum, Lanczos, Krylov, Eigenvalue, Eigenvector,
  Hermitian, LANSO, Lattice.
\end{abstract}
\date{\small{\it Version of \today}}
\maketitle
%
%To do
%-----
% - Previous and alternative approaches: Ritz, Arnoldi[DONE], PRIMME.
% - Costs of explicit Gram--Schmidt versus selective reorthogonalization.
% - Accuracy choices, i.e., \(\epsilon\) need not be ULP.
% - Brief discussion of applications (projecting low modes of kernel Dirac
% - operator to accelerate computation of overlap operator and its inverse, low
%   mode averaging, etc.)[NEEDS EXPANDING]
% - using previous lambda as shift for QR

\section{Introduction} \label{sec:intro}

\subsection{Motivation}

The problem of computing part of the spectrum of a large Hermitian matrix is
common to many areas of computational science, but the particular application
that motivated this work is the computation of the Neuberger operator for
lattice QCD (Quantum Chromodynamics being the quantum field theory of the
strong nuclear force).  This requires us to evaluate the \emphname{signum}
function of the ``Hermitian Dirac operator'' \(\gamma_5D\) corresponding to
some discrete lattice Dirac operator \(D\), which is defined by diagonalizing
this matrix and taking the \emphname{signum} (\(\pm1\)) of each of its
eigenvalues.  It is far too expensive to carry out the full diagonalization,
so we use a Zolotarev rational approximation for the \emphname{signum}
function as this can be evaluated just using matrix addition, multiplication,
and inversion by using a multi-shift solver for its stable partial fraction
expansion~\cite{Kennedy:2006ax}.  The approximation is expensive for
eigenvalues of \(\gamma_5 D\) that are very close to zero, and as there are
only a relatively small number of these we want to deflate them and take their
sign explicitly.  For this reason we need to compute the part of the spectrum
of \(\gamma_5 D\) around zero.

\subsection{Outline}

We begin surveying some basic properties of symmetric matrices in order to
introduce the notation used throughout the paper.  A pedagogical review of
simple eigensolver methods then follows, which leads on to the derivation of
the \Lanczos\ method with an explanation of the problems associated with it
when using finite-precision floating point arithmetic.  An overview of the
\emphname{LANSO} algorithm of Parlett and Scott~\cite{Parlett:1979:LAS} is
introduced which forms the starting point for the work described here.  The
goal of the algorithm we introduce in this paper is not to find the full
spectrum of a large Hermitian matrix, but to find that part of the spectrum
lying within some specified range.  For the application described in
\secref{sec:calcferm} its implementation in Chroma~\cite{Edwards:2004sx}
performs significantly better than the state--of--the--art Ritz
method~\cite{Kalkreuter:1995mm,Bunk:1996kt}.

\section{Hermitian Matrices and the Power Method}

\subsection{Basic Properties of Symmetric Matrices}

A matrix \(\ma A\) is \emphdef{Hermitian} (with respect to a sesquilinear
inner product) if \(\ma A = \ma A^\dagger\), which means \( (\vect u,\ma A
\vect v) = (\ma A^\dagger \vect u, \vect v) = (\ma A \vect u, \vect v) =
(\vect v,\ma A \vect u)^{*}, \) or equivalently \(\vect u^\dagger\cdot \ma A
\vect v = (\ma A^\dagger \vect u)^\dagger\cdot \vect v = (\ma A \vect
u)^\dagger\cdot \vect v = (\vect v^\dagger\cdot \ma A \vect u)^{*}.\) An
eigenvalue \(\lambda\) of \(\ma A\) satisfies \(\ma A \vect z=\lambda \vect
z\) where \(\vect z\neq 0\) is the corresponding eigenvector.  The eigenvalues
are real and the eigenvectors are orthogonal.  Any matrix can be reduced to
triangular form \(\ma T\) by a unitary (orthogonal)
transformation\footnote{This is \emphname{Schur normal form}, which follows
from the Cayley--Hamilton theorem that every matrix satisfies its
characteristic equation, and the fundamental theorem of algebra which states
that the characteristic polynomial \(p(\lambda) = \det(\ma A - \lambda)\) has
exactly \(N=\dim(\ma A)\) complex roots, counting multiplicity.} (change of
basis), \(\ma A = \ma Q\ma T\ma Q^{-1} = \ma Q\ma T\ma Q^\dagger\).  For \(\ma
A\) Hermitian \(\ma T^\dagger = (\ma Q^\dagger \ma A\ma Q)^\dagger = \ma
Q^\dagger \ma A^\dagger \ma Q = \ma Q^\dagger \ma A\ma Q = \ma T \) it follows
that \(\ma T\) is real and diagonal; thus \(\ma A \ma Q = \ma Q \ma T\) so the
columns of \(\ma Q\) furnish the orthonormal eigenvectors.

\subsection{Power Method} \label{sec:power}
   
In order to find eigenvalues and eigenvectors numerically one obvious approach
is the \emphdef{Power Method}.  An arbitrary starting vector can, in theory,
be expanded in the orthonormal eigenvector basis \(\{\vect z_j\}\), \(\vect
u_{0} = \sum_j \vect z_j (\vect z_j, \vect u_0)\).  The matrix \(\ma A\) is
applied to \(\vect u_0\) and the result normalized to get \(\vect u_1\), and
so forth: \(\vect u_{k+1} = \ma A \vect u_k/\|\ma A \vect u_k\|\), where the
norm is \(\|\vect x\| = \sqrt{(\vect x, \vect x)}\).  We then find that
\(\vect u_k \propto \lambda_1^k \vect z_1(\vect z_1, \vect u_0) + \sum_{j>1}
(\lambda_j/\lambda_1)^k \vect z_j(\vect z_j, \vect u_0) \), and as \(\lim_{k
  \to \infty} (\lambda_j/\lambda_1)^k = 0\) we find \(\lim_{k \to \infty}
\vect u_k = \vect z_1\) assuming \((\vect z_1,\vect u_0) \neq 0\), where we
label the eigenpairs such that \(|\lambda_1| > |\lambda_2| >
\cdots>|\lambda_N|\).  If the eigenvalue \(\lambda_1\) is degenerate then
\(\vect u_k\) converges to the eigenvector parallel to \(\vect u_0\).  The
rate of convergence is governed by \(\left|\lambda_2/ \lambda_1\right|^k =
e^{-k(\ln|\lambda_1|-\ln|\lambda_2|)}\).  If we shift the matrix \(\ma A\) by
a constant then we just shift its eigenvalues by the same constant and leave
the eigenvectors unchanged; however, such a shift \emph{does} change the rate
of convergence of the power method.

\section{Krylov Spaces and \lanczos\ Algorithm} \label{sec:kry}

\subsection{Krylov Spaces}

We consider a sequence of subspaces of increasing dimension \(n\) such that
the restriction of \(\ma A\) to them converges to \(\ma A\) as \(n\to\infty\).
For an \(N\times N\) matrix \(\ma A\), convergence will always occur because
the approximations equal \(\ma A\) for \(n\geq N\).  In many cases of
practical interest the matrix approximates some compact linear operator on an
\(\infty\)-dimensional Hilbert space, and we expect the convergence to be
governed by the properties of the underlying operator.

In practice we usually do not have an explicit matrix representation of the
large (sparse) matrix \(\ma A\), but we merely have some functional ``black
box'' representation that allows us to apply it to a vector in \( \R^N\).
Almost the only spaces we can construct from this are the \emphdef{Krylov
  spaces} \(\Krylov(\ma A, \vect u,n) = \Span(\vect u,\ma A \vect u,\ma A^2
\vect u,\ldots,\ma A^{n-1} \vect u)\) where \(\vect u\) is some more-or-less
arbitrary starting vector.  The only simple generalization is \emphdef{block
  Krylov spaces} where we start from more than one vector.

\subsection{Arnoldi Method}

The vectors \(\{\ma A^j \vect u\}\) do not form an orthonormal basis for the
Krylov space.  Furthermore, the corresponding unit vectors \(\ma A^j \vect
u/\|\ma A^j \vect u\|\) converge to the largest eigenvector of \(\ma A\), as
they are just successive iterates of the power method.  They therefore provide
a particularly \emph{bad} choice of basis for numerical computations.  It is
natural to construct a better orthonormal basis by deflation and
normalization, \[\vect q_1 = \vect u/\|\vect u\|, \qquad \vect u_{j+1} = \ma A
\vect q_j - \sum_{k=1}^j \vect q_k (\vect q_k,\ma A \vect q_j), \qquad \vect
q_{j+1} = \frac{\vect u_{j+1}}{\|\vect u_{j+1}\|};\] in other words the
Gram--Schmidt procedure.  This is called the \emphdef{Arnoldi method}.  We see
immediately that \((\vect q_{j+1},\ma A \vect q_j) = (\vect q_{j+1}, \vect
u_{j+1}) = \|\vect u_{j+1}\|\).  The \(n\times n\) matrix \(\ma Q\) whose
columns are\footnote{\(\vect e_j\) is a basis vector whose components are
  \([\vect e_j]_i=\delta_{ij}\).} \(\ma Q \vect e_j= \vect q_j\) therefore
furnishes an orthogonal projector \(\ma Q \ma Q^{\dagger} = \sum_{j=1}^n \vect
q_j \otimes \vect q_j^{\dagger}\) onto \(\Krylov(\ma A, \vect u, n)\).

The restriction of \(\ma A\) to the Krylov space is \emphdef{Hessenberg} by
construction: 
\[
  \ma H = \ma Q^\dagger \ma A \ma Q = \left(
  \begin{array}{cccccc}
    H_{1,1} & H_{1,2} & & H_{1,n-2} & H_{1,n-1} & H_{1,n} \\ 
    H_{2,1} & H_{2,2} & \cdots & H_{2,n-2} & H_{2,n-1} & H_{2,n} \\ 
    0 & H_{3,2} & & H_{3,n-2} & H_{3,n-1} & H_{3,n} \\ 
    \vdots & & \ddots & & \vdots & \\ 
    0 & 0 & & H_{n-1,n-2} & H_{n-1,n-1} & H_{n-1,n} \\ 
    0 & 0 & \cdots & 0 & H_{n,n-1} & H_{n,n}
  \end{array}\right).
\]

We can diagonalize this matrix using the \QR\ algorithm~\cite{Golub:1996} to
obtain \(\Theta=\ma S^\dagger \ma H \ma S\), where \(\Theta\) is the diagonal
matrix of \emphdef{Ritz values}, \(\Theta_{ij}=\theta_j\delta_{ij}\), and
\(\ma S\) the \(n\times n\) unitary (orthogonal) matrix whose columns are the
corresponding \emphdef{Ritz vectors} \(\vect s_j=\ma S \vect e_j\).  We may
hope that some of the Ritz values approximate the eigenvalues of \(\ma A\),
\(\theta_j\approx\lambda_{j'}\), and that some of the Ritz vectors approximate
its eigenvectors, \(\ma Q \ma S \vect e_j=\ma Q \vect s_j= \vect y_j\approx
\vect z_{j'}\), provided that the \emphdef{residual} $\ma R \defn \ma A \ma Q
- \ma Q \ma H$ is small, since \(\ma A(\ma Q \ma S) = (\ma Q \ma H + \ma R)
\ma S = (\ma Q \ma S)\Theta + \O(\|\ma R\|)\).

\subsection{\lanczos\ Algorithm}

We are interested the special case of the Arnoldi method for a Hermitian
matrix \(\ma A\), which means that the matrix \(\ma H\) is also Hermitian,
\(\ma H^\dagger = (\ma Q^\dagger \ma A \ma Q)^\dagger = \ma Q^\dagger \ma
A^\dagger \ma Q = \ma H\).  A matrix which is both Hessenberg and Hermitian is
\emphdef{tridiagonal}
\[
  \ma H = \ma Q^\dagger \ma A \ma Q = \left(
    \begin{array}{ccccccc}
      \alpha_1 & \beta_1 & 0 & & 0 & 0 & 0 \\
      \beta_1 & \alpha_2 & \beta_2 & \cdots & 0 & 0 & 0 \\
      0 & \beta_2 & \alpha_3 & & 0 & 0 & 0 \\
      & \vdots & & \ddots & & \vdots & \\
      0 & 0 & 0 & & \alpha_{n-2} & \beta_{n-2} & 0 \\
      0 & 0 & 0 & \cdots & \beta_{n-2} & \alpha_{n-1} & \beta_{n-1} \\
      0 & 0 & 0 & & 0 & \beta_{n-1} & \alpha_n      
    \end{array}\right),
\] 
where \(\beta_j=\|\vect u_{j+1}\|=(\vect q_{j+1},\ma A \vect q_j)\) and
\(\alpha_i=(\vect q_i,\ma A \vect q_i)\) are real.

We thus have a three-term recurrence relation 
\begin{equation} \label{eqn:threeterm}
  \ma A \vect q_j 
    = \beta_j \vect q_{j+1} + \alpha_j \vect q_j + \beta_{j-1} \vect q_{j-1};
\end{equation}
this defines the \emphdef{\lanczos\ algorithm}.  This greatly simplifies the
computation; not only is it easier to diagonalize a tridiagonal matrix using
the \QR\ algorithm, but also means that \(\ma A \vect q_j\) is automatically
(implicitly) orthogonal to all \(\vect q_i\) except for \(\vect q_{i-1}\),
\(\vect q_i\), and \(\vect q_{i+1}\).  Unfortunately, floating-point
arithmetic does not respect implicit orthogonality.

\subsection{Loss of orthogonality among the \Lanczos\ vectors}

As noted in the previous section, with a basic implementation of the
\Lanczos\ algorithm, orthogonality amongst the \Lanczos\ vectors is lost due
to rounding errors.  The most obvious indication of this loss of orthogonality
is the appearance of spurious copies of eigenvectors.  It is interesting to
store the \Lanczos\ vectors to measure the loss of orthogonality directly, as
it allows us to see where the loss of orthogonality occurs.  The results are
as expected: with the basic \Lanczos\ algorithm (\ie one with no
reorthogonalization) the orthogonality of a \Lanczos\ vector with respect to
those calculated more than two steps previously is only implicit;
consequently, as rounding errors inevitably bring back components of the early
\Lanczos\ vectors, there is nothing to suppress these components.  They
therefore grow in an unrestrained manner until eventually orthogonality
between the most recent \Lanczos\ vector and those calculated early on in the
procedure is completely lost.  This was demonstrated in~\cite{simon:1984}
where\footnote{The ``unit of least precision'', \(\ulp\), is the smallest
  number such that \(1\oplus\ulp\neq1\) in floating-point arithmetic, it is
  approximately \(10^{-7}\) for single precision and \(10^{-14}\) for double
  precision.} \(\log_{10}(\vect q^*_j\vect q_k/\ulp)\) is displayed as a
symmetric array of numbers with small values representing mutually orthogonal
vectors and large ones representing pairs of vectors with a large overlap.

\begin{figure}[!ht]
  \begin{center}
    \epspdffile{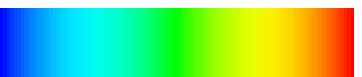}
  \end{center}
  \vspace{-0.5cm}
  \caption*{\(\uparrow \hspace{10.7cm} \uparrow\)}
  \caption*{\((\vect q_i, \vect q_j) = 0\) \hspace{7.5cm} \((\vect q_i, \vect
    q_j) = 1\)}
  \vspace{0.5cm}
  \begin{center}
    \epspdffile{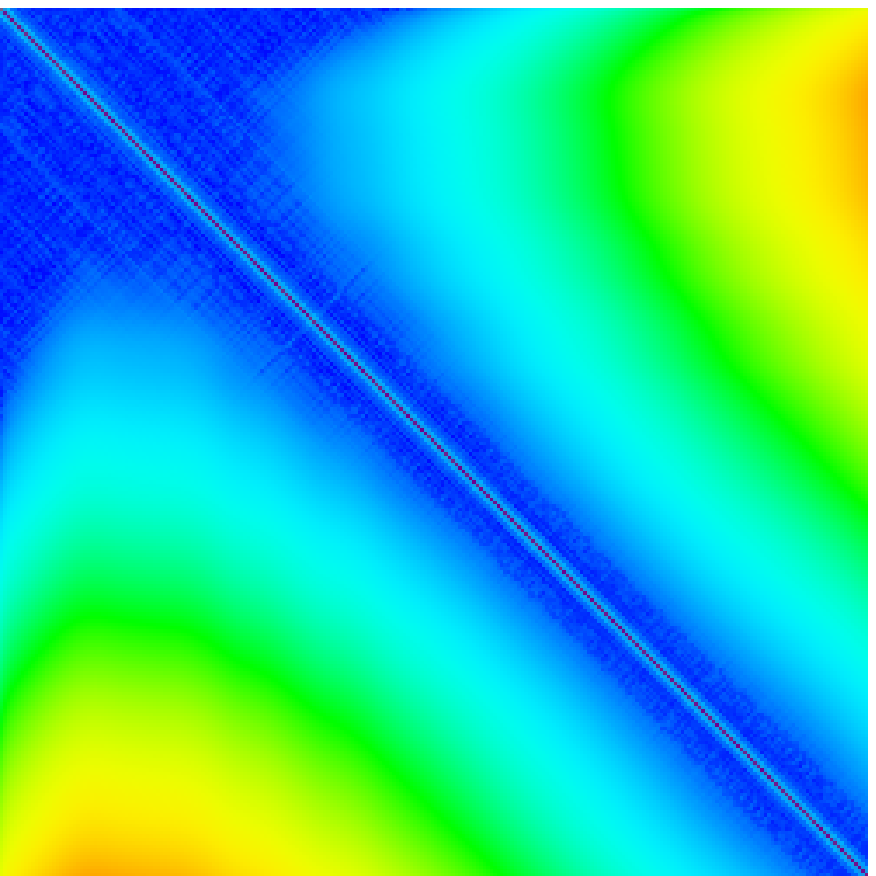}
  \end{center}
  \caption{The orthogonality of the \Lanczos\ vectors without
    reorthogonalization.  Increasing \Lanczos\ iterations, \(i\) and \(j\) are
    shown in the \(\downarrow\) and \(\rightarrow\) directions.}
  \label{fig:qq2}
\end{figure}

For larger systems we can view this as a colour map, an example of which is
shown in \figref{fig:qq2}.  Large values are represented at the red end of the
spectrum and small values at the blue end.  Thus the diagonal is shown in red
(representing \(\|q_j\| = 1\)) and mutually orthogonal vectors are shown
in~blue.

We can see clearly that with no reorthogonalization, after sufficient steps
the new \Lanczos\ vectors lose orthogonality to the early ones.  Note that
\(\vect q_j\) loses orthogonality with the very earliest \Lanczos\ vectors to
a lesser extent compared with those which occur after a few steps.  This is to
be expected as the initial random starting vector \(\vect q_1\) will in
general not contain large components of any particular eigenvector.  However,
after a few steps the \Lanczos\ vectors will start to contain large components
of the dominant eigenvectors according to the argument given in
\secref{sec:power} for the power method, and it is precisely these dominant
eigenvectors that will grow from rounding errors and so reappear in~\(\vect
q_j\).

\subsection{Degenerate Eigenspaces and Restarting}

In exact arithmetic only one eigenvector will be found for each distinct
eigenvalue: if an eigenvalue is degenerate then this vector will be the
projection of the initial vector onto its eigenspace.  In floating-point
arithmetic rounding errors will eventually cause the other eigenvectors to
appear; this will take longer in higher-precision arithmetic.  This may
perhaps be viewed as a case where using floating-point arithmetic is an
advantage.  Such degenerate eigenvectors can also be found by restarting the
\lanczos\ algorithm with a new initial vector and deflating with respect to
the previously known good eigenvectors.  This can be repeated until no more
degenerate eigenvectors are found.  Presumably a block version of the
algorithm could be used too, but the choice of block size is not obvious if
the maximum degeneracy is not known \emph{a priori}.  A cluster of nearby
eigenvalues behaves just like a degenerate subspace until sufficient accuracy
to resolve the eigenvalues has been attained.

\section{Selective Reorthogonalization} \label{sec:so}

We will deem a Ritz vector \(\vect y_j \in \Krylov(\ma A, \vect u, n)\), where
to be ``good'' if it lies within the Krylov subspace \(\Krylov(\ma A, \vect u,
n')\) with \(n'<n\), that is if \((\vect y_j, \vect q_k) = (\ma Q \vect s_j,
\ma Q \vect e_k) = (\vect s_j, \vect e_k)\approx0\) for~\(k>n'\); eigenvalues
that are not good will be called ``bad''.  Paige \cite{280490} has shown that
the loss of implicit orthogonality occurs primarily in the direction of good
Ritz vectors.  This is not surprising: if \(\vect q_{n'+1}\) and \(\vect
q_{n'+2}\) are orthogonal to an eigenvector \(\vect z\) of \(\ma A\) with
eigenvalue \(\lambda\) then all future \lanczos\ vectors will also be
orthogonal to \(\vect z\) in exact arithmetic.  We may prove this by
induction: assume \((\vect z, \vect q_k) = (\vect z, \vect q_{k+1}) = 0\) for
some \(k>n'\), then
\begin{eqnarray*}
  (\vect z, \ma A \vect q_{k+1}) & = & (\ma A \vect z, \vect q_{k+1})
  = \lambda (\vect z, \vect q_{k+1}) = 0 \\
  & = & (\vect z, \beta_{k+1} \vect q_{k+2} + \alpha_{k+1}\vect q_{k+1} +
    \beta_k \vect q_k) 
  = \beta_{k+1}(\vect z, \vect q_{k+2}),
\end{eqnarray*}
hence \((\vect z, \vect q_{k+2}) = 0\) unless the \Lanczos\ process terminates
because~\(\beta_{k+1} = 0\).  Concomitantly, any rounding errors that appear
in the computation of \(\vect q_j\) for \(j>n'+2\) with a component parallel
to \(\vect z\) will not be suppressed by orthogonalization to the previous two
\lanczos\ vectors; moreover, this component will grow as \(|\lambda /
\lambda'|^k\) where \(\lambda'\) is the largest ``bad'' eigenvalue.

It therefore suffices to orthogonalize the current \lanczos\ vectors \(\vect
q_n\) and \(\vect q_{n+1}\) explicitly with respect to good eigenvectors
sufficiently frequently.  This is much cheaper than explicitly
orthogonalizing with respect to all the previous \lanczos\ vectors at each
step as in the Arnoldi method.

\subsection{LANSO}

How often do we need to carry out this reorthogonalization? As rounding errors
are of order \(\ulp\) it seems reasonable to choose to do so when the loss of
orthogonality has accumulated to be of \(\O(\sqrt{\ulp})\).  We therefore
choose to orthogonalize \(\vect q_{n'}\) and \(\vect q_{n'+1}\) with respect
to a good Ritz vector \(\vect y\) when \((\vect y, \vect
q_{n'})>\sqrt{\ulp}\).  In their \emphname{LANSO} algorithm Parlett and
Scott~\cite{Parlett:1979:LAS} introduce two bounds,
\begin{enumerate}
\item The \(\tau\) bound, \(\tau_{ij}\geq|(\vect y_i, \vect q_j)|\), that is
  used to trigger reorthogonalization with respect to \(\vect y_i\).  This
  bound is computed cheaply by a three-term scalar recurrence.
\item The \(\kappa\) bound, \(\kappa\geq\|\ma Q^\dagger \ma Q-1\|\), that is
  used to trigger a ``pause'', namely a search for new good eigenvectors by
  running the \QR\ algorithm, followed by a reorthogonalization of the last
  two \Lanczos\ vectors with respect to all good eigenvectors.  This is
  computed by a more complicated scalar recurrence.
\end{enumerate}

\subsubsection[Monitoring the kappa and tau bounds]{Monitoring the \(\kappa\)
  and \(\tau\) bounds} \label{sec:mon-bound}

\begin{figure}[!ht]
  {\epsfxsize=.7\textwidth
  \begin{center}
    \epspdffile{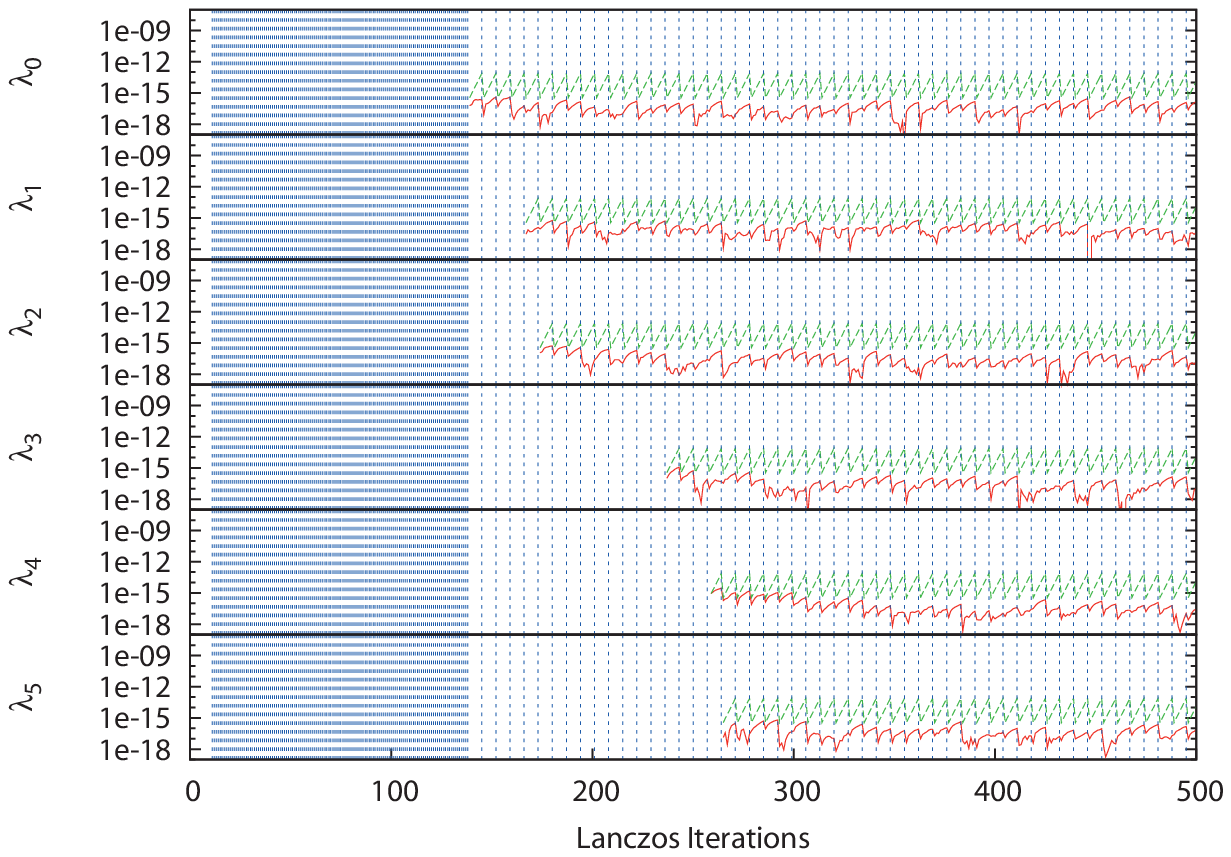}
  \end{center}
  \begin{center}
    \epspdffile{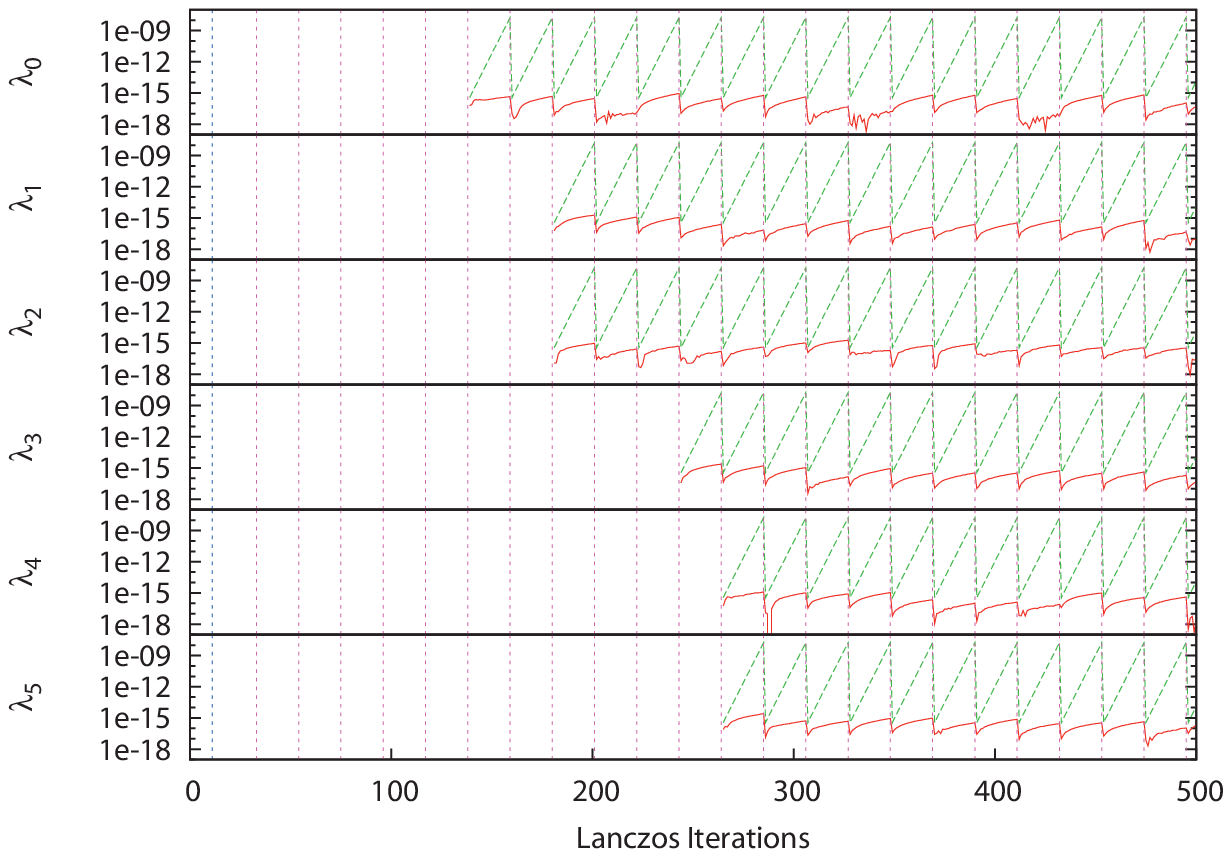}
  \end{center}}
  \caption{The upper figure plots the values of \(|(\vect y_k, \vect q_j)|\)
    in red and its bound \(\tau_{kj}\) in green for each \(\vect y_k\) for
    several different Ritz values \(\lambda_k\).  The blue vertical dotted
    lines show the points at which a diagonalization of \(\ma H\) is triggered
    by the \(\kappa\) bound.  The lower figure is similar but here the red
    vertical lines show the \(\sigma\) bound being used to trigger a full
    diagonalization.  In this example the \(\kappa\) bound was used to trigger
    the very first pause, but this is not needed: \(\sigma\) could be used
    from the beginning.}
  \label{fig:bounds-kappa}
\end{figure}

The success of the LANSO method hinges on the ability of \(\kappa_j\) and
\(\tau_{kj}\) to bound well enough the actual values \(\|1 - \ma Q^* \ma Q\|\)
and \((\vect y_k, \vect q_j)\) respectively for any \Lanczos\ step \(j\) and
all good eigenvectors \(\vect y_k\) calculated thus far.  For our relatively
small test cases we can store all the \Lanczos\ vectors which make up \(\ma
Q\), and all the known good eigenvectors.  This enables us to calculate the
values of \(\|1 - \ma Q^\dagger\ma Q\|\) and \((\vect y_k, \vect q_j)\) to
compare with these bounds.  This information is plotted in
\figref{fig:bounds-kappa} where the \(\tau\) bound is plotted together with
the value which it is supposed to bound.  The points at which the \(\kappa\)
bound triggers a pause are also shown.  This figure reveals a number of
features.  Firstly, \(\tau_{kj} > |(\vect y_k, \vect q_j)|\) as required.
However, the bounds appear rather pessimistic: the \(\kappa\) bound exceeds
the tolerance \(\sqrt{\ulp}\) and triggers a pause (recall this entails
the calculation of the spectrum of the tridiagonal matrix) far more frequently
than needed, and the \(\tau\) bound is often many orders of magnitude larger
than the quantity it is bounding; however, the reorthogonalization triggered
by this is relatively inexpensive.  Due to the frequent triggering by the
\(\kappa\) bound, in practice it is the \(\kappa\) bound and never the
individual \(\tau\) bounds which triggers reorthogonalization.

\subsection{New Algorithm}

In our application, as in many others, we do not need to find all the
eigenpairs: it suffices to find those in a pre-specified region \(\Sigma\) of
the spectrum.  We only need search for eigenvalues in \(\Sigma\) and
selectively reorthogonalize \Lanczos\ vectors with respect to them; we are not
interested if duplicate eigenvectors occur outside~\(\Sigma\).  In passing, we
note that it is easy to restrict the \QR\ iteration to search in the region by
a judicious choice of shifts (see \secref{sec:QR}).

Our algorithm replaces both LANSO bounds with a bound \(\sigma\) that is a
generalization of the \(\tau\) bound.  \(\sigma\) bounds the loss of
orthogonality of a \Lanczos\ vector \(\vect q_j\) with respect to \emph{any}
good Ritz vector \(\vect y\) within~\(\Sigma\), even if \(\vect y\) is not
explicitly known.  We shall require that \(\sigma_j \ge \max_{k:\theta_k \in
  \Sigma}|(\vect y_k, \vect q_j)| ,\) where the maximum is taken over all good
Ritz pairs in~\(\Sigma\).

\(\sigma\) is calculated via a three term recurrence relation closely related
to that for the \(\tau\) bound.  We consider the propagation and amplification
of the lack of orthogonality of the good Ritz vectors with current
\Lanczos\ vectors and ignore other inconsequential rounding errors as
in~\cite{Parlett:1979:LAS}.  Taking the inner product of~\eqref{eqn:threeterm}
with \(\vect y_k\) gives
\begin{equation} \label{eqn:yaq1}
  (\vect y_k,\mtx A \vect q_j)
    - (\vect y_k,\vect q_{j-1}) \beta_{j-1}
    - (\vect y_k,\vect q_j) \alpha_j
    - (\vect y_k,\vect q_{j+1}) \beta_j = 0.
\end{equation}
If \(\vect y_k = \mtx Q \vect s_k\) is a good Ritz vector within \(\Sigma\),
where \((\theta_k,\vect s_k)\) is a Ritz pair (\(\mtx H \vect s_k = \vect s_k
\theta_k\)) then
\begin{equation} \label{eqn:yaq2}
  (\vect y_k,\mtx A\vect q_j) = (\mtx A \vect y_k,\vect q_j)
  = (\mtx A \mtx Q \vect s_k,\vect q_j)
  = (\mtx Q \mtx H \vect s_k,\vect q_j) + (\mtx R \vect s_k,\vect q_j)
  = (\vect y_k,\vect q_j) \theta_k ,
\end{equation}
as the residual is orthogonal to the Krylov space, \(\mtx Q^{\dag} \mtx R =
\mtx Q^{\dag} \mtx A \mtx Q - \mtx Q^{\dag} \mtx Q \mtx H = 0\).  From
\eqref{eqn:yaq1} and \eqref{eqn:yaq2} we obtain
\[
  (\vect y_k,\vect q_{j+1}) \beta_j
  = (\vect y_k,\vect q_j) (\theta_k-\alpha_j)
    - (\vect y_k,\vect q_{j-1}) \beta_{j-1}  .
\]
We assume by induction that \(\sigma_i \ge |(\vect y_k,\vect q_i)| \; \forall
k: \theta_k \in \Sigma, \forall i \le j\), hence
\begin{eqnarray*}
  |(\vect y_k,\vect q_{j+1})|\,|\beta_j|
  &\leq& |(\vect y_k,\vect q_j)|\,|\theta_k-\alpha_j|
    + |(\vect y_k,\vect q_{j-1})|\,|\beta_{j-1}| \\
  &\leq& \max_{\theta \in \Sigma} \sigma_j |\theta-\alpha_j| + \sigma_{j-1}
    |\beta_{j-1}| ; 
\end{eqnarray*}
so we may define
\begin{displaymath}
  \sigma_{j+1} = \frac{\displaystyle \max_{\theta\in\Sigma}|\theta-\alpha_j|
    \sigma_j + |\beta_{j-1}|\sigma_{j-1}}{|\beta_j|},
\end{displaymath}
where the ``initial values'' \(\sigma_{t-1}=\O(\ulp)\) and \(\sigma_t =
\O(\ulp)\) correspond to the lack of orthogonality after selectively
orthogonalizing a good Ritz vector using a finite precision arithmetic
implementation of Gram--Schmidt, \(t\) being the last iteration at which the
algorithm was paused to search for new Ritz pairs.  We shall not give a
detailed analysis of this algorithm here but it is very similar for that for
LANSO given in~\cite{280490}.

We are interested in applying our algorithm to low density interior regions of
the spectrum.  The algorithm is surprisingly effective as we find such
interior eigenvalues converge rapidly in a manner reminiscent of extremal
eigenvalues.  The reason why eigenvalues in low-density regions are so well
represented in the Krylov space is explained in~\cite{Johnson:2011py}.

% Our method only selectively orthogonalizes good eigenvectors within
% \(\Sigma\), so eigenvectors not lying within \(\Sigma\) will recur.  This
% does not matter as we are not interested in such eigenvectors.  Furthermore
% they will only recur infrequently if \(K(\ma A) \epsilon\) is very small.  We
% may expect the recurrence frequency to be \(-\ln(K(\ma A)) /
% \ln(\epsilon\)).

\subsubsection{Results of the new algorithm}

The lower panel of \figref{fig:bounds-kappa} shows the effect of using
\(\sigma\) to trigger a pause.  We see immediately that when using the
\(\sigma\) procedure, diagonalization of \(\ma H \) is performed far less
frequently than is the case when using the \KAPPA\ procedure.

\section{Calculating low eigenvalues of the Fermion matrix} \label{sec:calcferm}

The \Lanczos\ method itself can be used to diagonalize only Hermitian
matrices, but the matrices are not required to be positive definite.  The
Wilson--Dirac fermion matrix \(\ma D\) is not Hermitian, but we can exploit
the fact that our matrix is ``\(\gamma_5\)-Hermitian'', \(\gamma_5 \ma D
\gamma_5 = \ma D^\dag\), where \(\gamma_5\) is a product of the four Hermitian
gamma matrices \(\gamma_5 = \gamma_1 \gamma_2\gamma_3\gamma_4\) which satisfy
the anticommutation relations \(\{\gamma_\mu,\gamma_\nu\} =
2\delta_{\mu\nu}\).  This allows us to construct the Hermitian matrix
\(\gamma_5 \ma D\).

We are interested in the eigenvalues close to gaps in the spectrum, and for
\(\gamma_5 \ma D\) there is such a gap around zero.  These eigenvalues map to
extremal eigenvalues of \(\ma D^{\dagger} \ma D = (\gamma_5 \ma D)^2\), but if
we use \(\ma D^{\dag}\ma D\) then we have to consider the extra work involved
in resolving the sign of the corresponding eigenvalues of \(\gamma_5 \ma D\).
This also involves dealing with any mixing which takes place due to the near
degeneracy of the approximate eigenvalues, since eigenvalues \(\lambda^2\) of
\(\ma D^{\dagger} \ma D\) might be mixtures of eigenvalues of \(\gamma_5 \ma
D\) near either \(\pm \lambda\).

\begin{figure}[!ht]
  \begin{center}
    \epspdffile{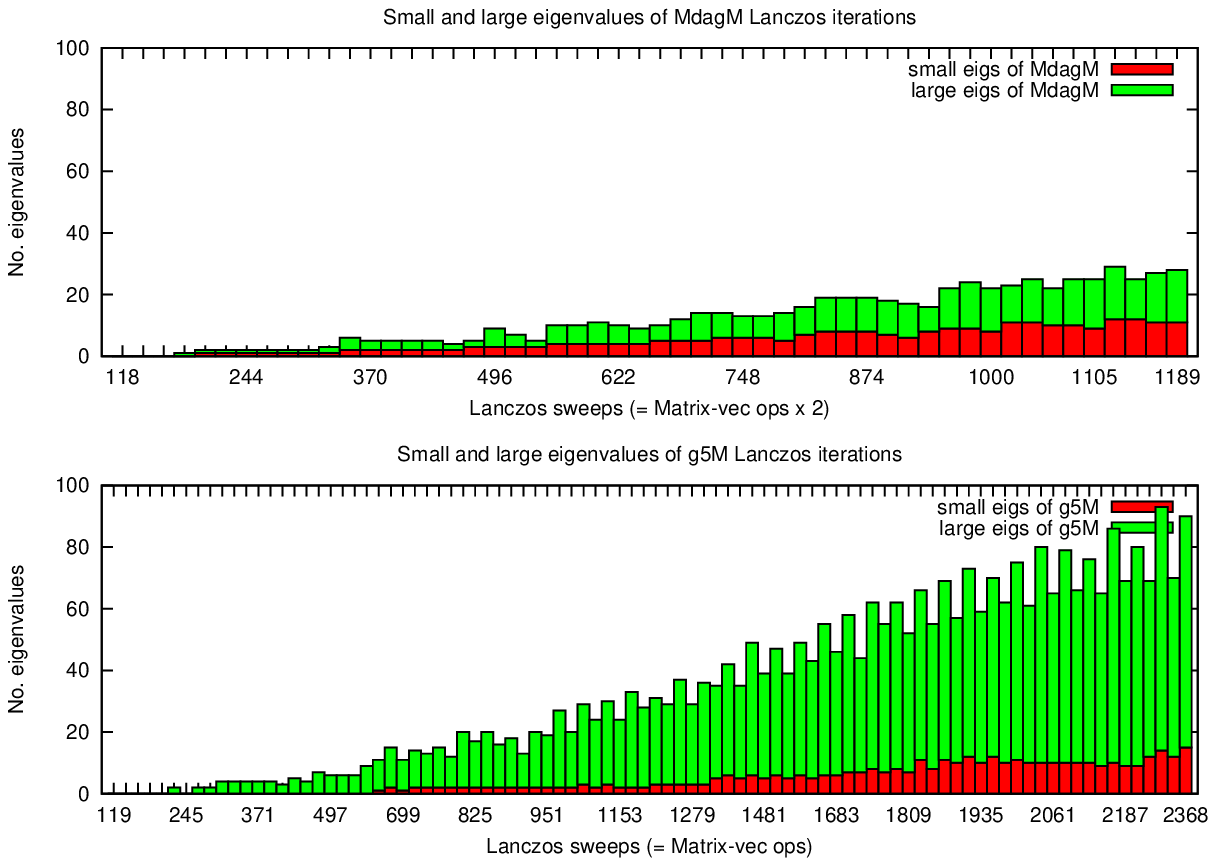}
  \end{center}
  \caption{The number of large and small magnitude eigenvalues of \(\ma H =
    \gamma_5 \ma M\) found as a function of the number of \Lanczos\ steps
    (dimension of the Krylov space).  The horizontal axes are scaled to same
    number of \(\ma\gamma_5D\) applications.}
  \label{fig:small-large}
\end{figure}

When using the \Lanczos\ method we see eigenvalues of both large and small
magnitude being resolved first giving two regions in the case of \( \ma
D^{\dag}\ma D \) (corresponding to both large and small eigenvalues), and four
regions in the case of \(\gamma_5 \ma D\) (corresponding to large and small
eigenvalues both positive and negative).  \figref{fig:small-large} shows a
bar chart of the relative number of small and large converged eigenvalues
(regardless of sign) determined at each pause for both \( \gamma_5 \ma D\) and
\(\ma D^{\dag}\ma D\).

The fact that \figref{fig:small-large} shows many more large eigenvalues being
resolved than small ones gives good motivation for our earlier assertion that
we should only look for eigenvalues within the region of interest.  If we
were to find, construct, and reorthogonalize with respect to all converged
eigenvalues at a given \Lanczos\ step most of the time would be spent
preserving the orthogonality of regions we are not interested in.

\begin{figure}[!ht]
  \begin{center}
    \epspdffile{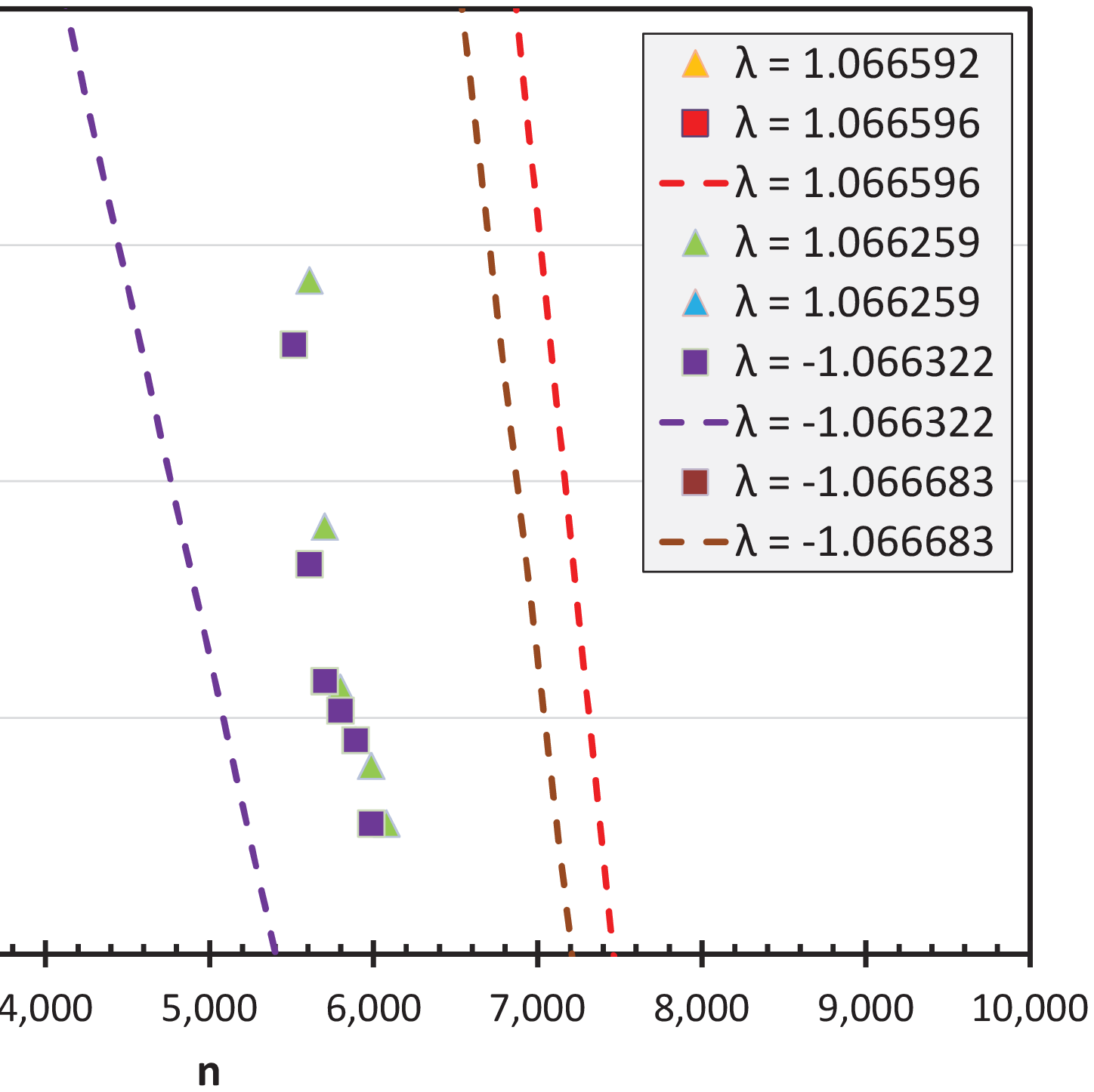}
  \end{center}
  \caption{Graph showing the error in eigenvalue estimates as a function of
    iteration number \(n\) (Krylov subspace dimension).  We compare this with
    the theoretical bounds indicated by the dashed lines.  The error is
    determined by taking the absolute value of the difference between the
    measured Ritz values and the nearest eigenvalue (approximated by the most
    accurate Ritz value we obtain at the end of the run).  The eigenvalue used
    for a given Ritz value is indicated by different symbols as indicated in
    the legend.  When the error is large this association is somewhat
    arbitrary but it is unambiguous for the range of errors shown in this
    graph.  The lines correspond to the bounds obtained using the results
    of~\cite{Johnson:2011py}, again using the spectrum as approximated using
    the most accurate Ritz values.  We see that the purple squares (\(\lambda
    = -1.066322\)) and green triangles (\(\lambda = 1.066259\)) seem to
    correspond to two orthogonal eigenvectors belonging to degenerate (or very
    nearly degenerate) eigenvalues.  If they were actually degenerate then the
    second eigenvector would be a fortuitous consequence of rounding error.
    The agreement between the observed rate of convergence and the theoretical
    bounds is quite satisfactory.}
  \label{fig:conv}
\end{figure}

As stated earlier, we are interested in the eigenvalues which are close to a
gap in the eigenspectrum around zero.  The convergence rates for extremal
eigevalues, \ie those at either end of the spectrum, are well understood
following the work of Kaniel~\cite{Kaniel:1966}, Paige~\cite{Paige:1971} and
Saad~\cite{Saad:1980}.  This explains why, in the case of \(\ma D^{\dagger}
\ma D\) where all the eigenvalues are positive, we see the largest and
smallest eigenvalues converge quickly compared with interior ones.  In the
case of the matrix \(\gamma_5 \ma D\) we see the eigenvalues smallest in
magnitude converging quickly.  These eigenvalues are not at the extremes of
the spectrum but are close to a relatively large void in the spectrum around
zero.  The convergence rates for such ``interior'' eigenvalues is explained
in~\cite{Johnson:2011py} where we consider the Kaniel--Paige--Saad bounds
applied to the shifted and squared matrix (in this case the optimal shift is
zero).  \figref{fig:conv} shows a comparison of our theoretical bounds with
the errors found when finding the eigenvalues close to a gap in the
eigenspectrum of the Fermion matrix.

\section{Implementation details}

All results obtained here have been obtained using the
Chroma~\cite{Edwards:2004sx} package running on 4,096 cores of the UK National
Supercomputing Service HECToR~\cite{hector}, after a prototype code was
initially implemented in Maple~\cite{maple}.  The Chroma implementation
consists of highly optimized parallel linear algebra routines specifically
written for lattice QCD, thus we can assume that matrix-vector products,
inner-products and general manipulation of vectors and matrices are already
optimized.  Here we seek to minimize the number of calls to these operations
but not to optimize them further.  However, we do give some consideration here
to patterns of access to large vectors stored in memory, particularly when
constructing eigenvectors, and we also consider some optimization of the
currently serial diagonalization of the tridiagonal matrix \(\ma H\) using the
\QR\ method.

\subsection{Constructing eigenvectors}

Following each application of the \QR\ method, we need to calculate the
vectors \(\vect y_i = \ma Q \vect s_i,\) where \(\vect s_i\) are the columns
of \(\ma S\), \ie the Ritz vectors.  This means that each good Ritz vector
\(\vect y_i\) is constructed as a linear combination of \Lanczos\ vectors.
The most straightforward method for constructing each eigenvector is via a
simple loop as follows
\begin{verbatim}
  DO i = 1 to # good Ritz vectors
    DO j = 1 to # Lanczos vectors
        y[i] = y[i] + q[j] * S[j,i]
    END DO
  END DO
\end{verbatim}
where the number of good Ritz vectors is expected to be much smaller than the
number of \Lanczos\ vectors.

However, this may not be the most efficient ordering.  After many
\Lanczos\ iterations we will have a large number of \Lanczos\ vectors and they
may not all be available in fast memory.  We therefore need to ensure that
once a \Lanczos\ vector is retrieved from memory we make the most efficient
use of it, reducing the need for multiple loads and stores of the vector to
and from memory.  It may even be that we cannot store all of the
\Lanczos\ vectors, and need to reconstruct them on the fly.  It therefore
makes sense to access (or reconstruct) each \Lanczos\ vector in turn and build
up the good Ritz vectors together, by interchanging the order of the loops
\begin{verbatim}
  DO j = 1 to # Lanczos vectors
    Recalculate/access q[j]
    DO i = 1 to # good Ritz vectors
      y[i] = y[i] + q[j] * S[j,i]
    END DO
  END DO
\end{verbatim}
In both cases the Ritz vectors \(\vect y_i\) are accessed and updated within
the inner loop but the second method should result in fewer accesses to the
\Lanczos\ vectors, \(\vect q_j\).  Experiments show an average speed-up of
approximately 50\% in this case.  %COULD SHOW GRAPH HERE

There are some further interesting architecture-dependent trade-offs that
could be investigated.  Depending on the amount of memory available and the
memory bandwidth we can choose between
\begin{enumerate}
\item Storing the \lanczos\ vectors in main memory (DRAM);
\item Storing the \lanczos\ vector in secondary storage (disk or Flash RAM);
\item Recomputing the \lanczos\ vectors at each pause.  This minimizes off-chip
  data transfer, and is ``embarrassingly parallel'' up to a few global sum
  operations (for inner products and norms).
\end{enumerate}
A full investigation of these options has not been performed here.

\subsection[Diagonalization of H: QR]{Diagonalization of \(\ma H\):
  QR} \label{sec:QR}

We need to pause the \Lanczos\ process periodically to determine the
eigenspectrum of the tridiagonal matrix \(\ma H\).  This can be achieved
efficiently using the iterative implicit \QR\ algorithm~\cite{Golub:1996} with
suitable shifts.

Many implementations of implicit \QR\ methods exist.  The results here were
obtained using Lapack~\cite{Anderson:1990:LPL:110382.110385} routines built on
top of BLAS~\cite{blas}, accelerated using the ACML library~\cite{acml}.  The
DSTEV Lapack routine could be used to determine all the eigenvalues, and
optionally all eigenvectors, of a symmetric tridiagonal matrix.  This works
well for our needs; however, we are only interested in eigenvalues from within
a region~\(\Sigma\), which can give a significant performance benefit.  We are
better off employing a routine such as DSTEVX which finds eigenvalues only
within a specified interval.  In the case where \(\Sigma\) is a non-contiguous
range, this may result in the routine being called several times, once for
each range, or the algorithm could be rewritten to work with a disjoint range.
We could also make use of previously known good eigenvalues as shifts, but
this has not been implemented.

\section{Results}

\begin{figure}[!ht]
  \begin{center}
    \epspdffile{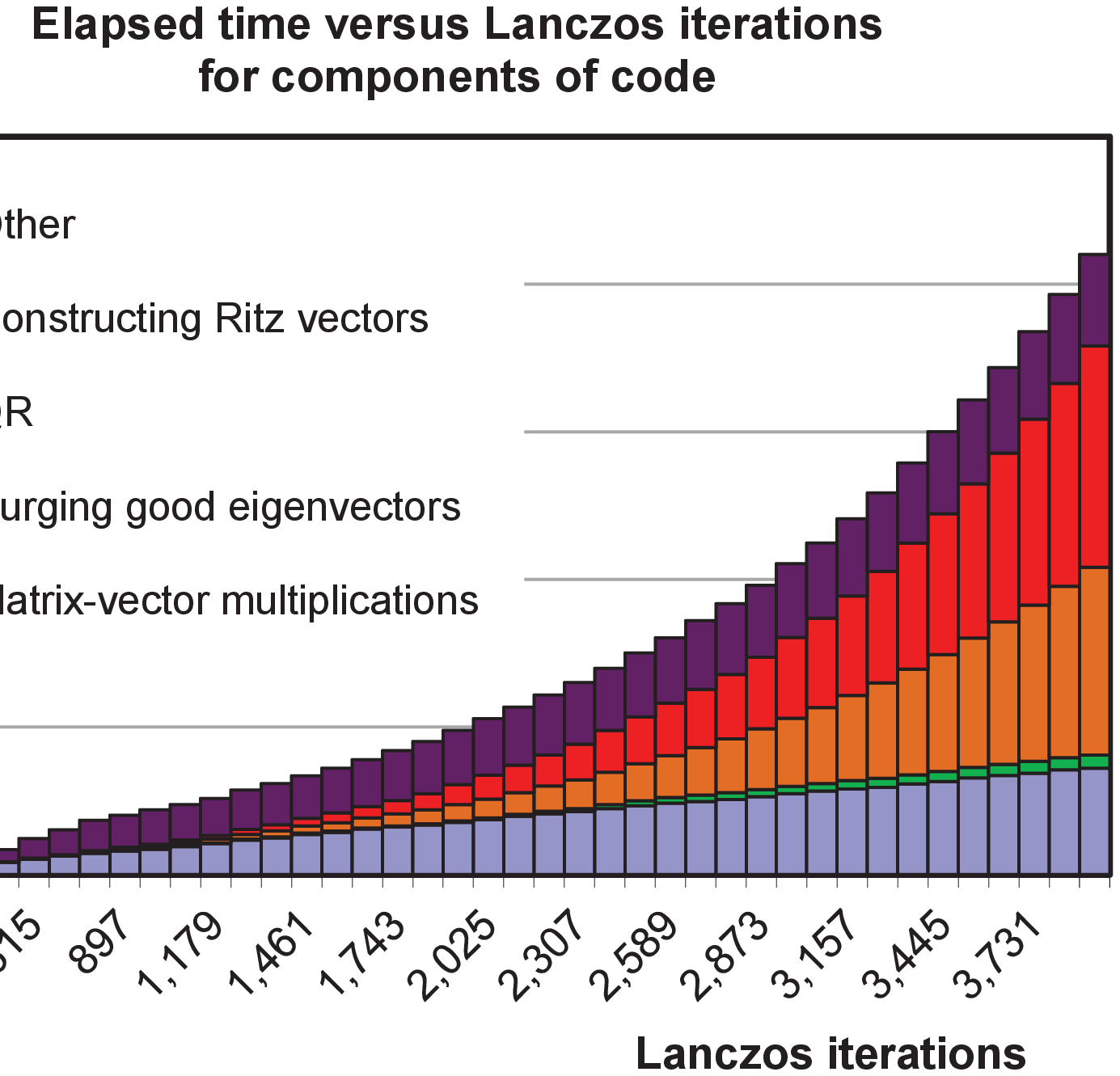}
  \end{center}
  \caption{Breakdown of time spent in various parts of the new algorithm
    versus \Lanczos\ iteration for \(\gamma_5 \ma D\) on a \(24^3 \times 48\)
    lattice with 12 degrees of freedom per lattice site (\ie\(N=7,962,624\)) on
    4,096 cores of a Cray~XT4.  ``Constructing Ritz vectors'' means computing
    \(\vect y = \ma Q \vect s\), and ``Purging good eigenvectors'' means
    reorthogonalising the last two \Lanczos\ vectors with all known good Ritz
    vectors. The \(x\)-axis shows the iteration numbers at which the algorithm
    is paused. The frequency of pauses is such that the \(x\)-axis scale is
    approximately linear.}
  \label{fig:breakdown4096}
\end{figure}

\figref{fig:breakdown4096} shows a breakdown of the various components of the
algorithm when running on the largest processor count attempted (4,096) for
our new algorithm applied to \(\gamma_5 \ma D\).  We find that with our
implementation the most expensive operation is the creation of the
eigenvectors of \(\gamma_5 \ma D\) followed by the application of the
\QR\ method, which is why we wish to create as few eigenvectors as possible.
It may also be desirable to implement a faster (e.g., parallel) \QR\ method as
the number of eigenvalues required becomes larger.

\begin{figure}[!ht]
  \begin{center}
    \epspdffile{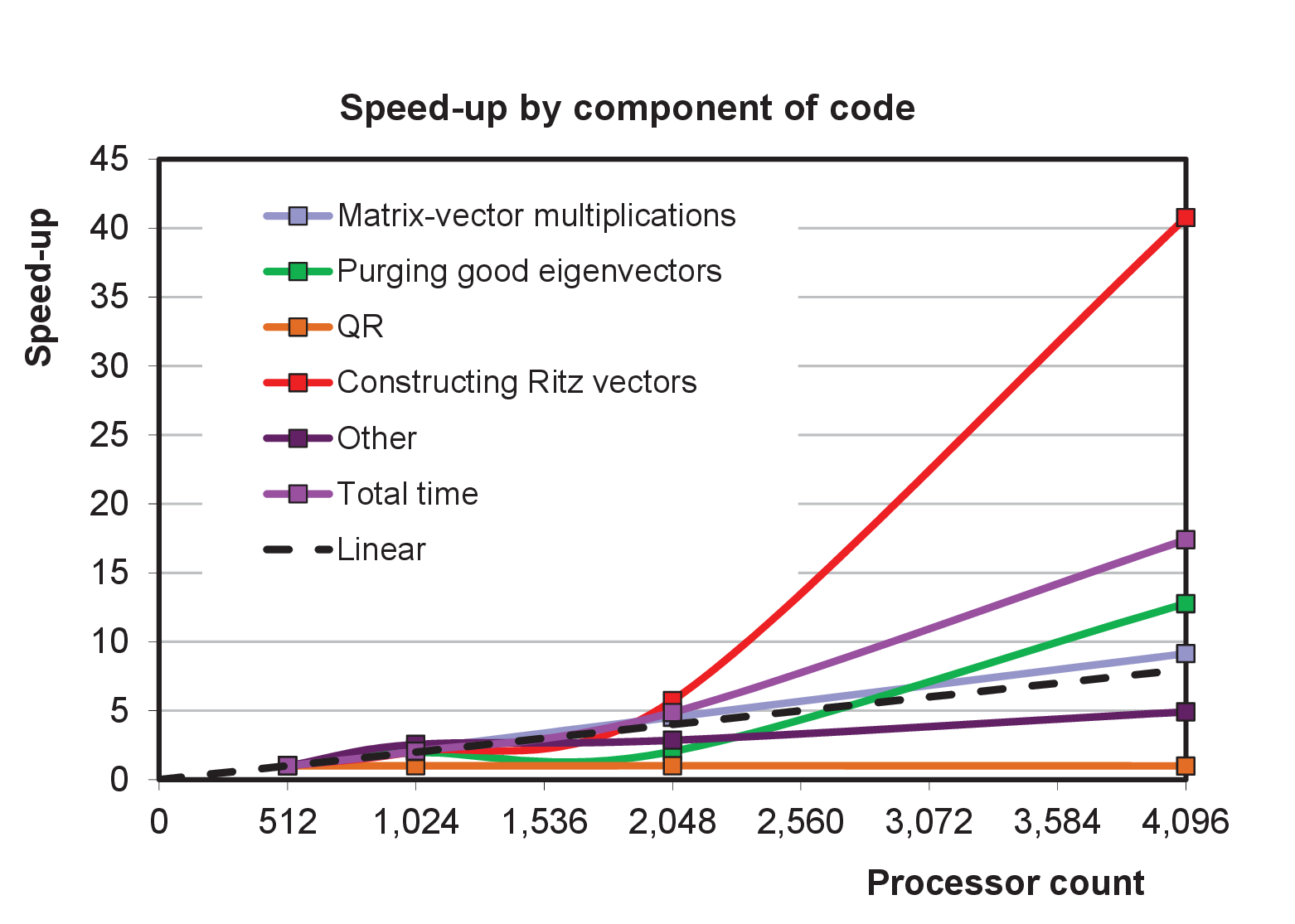}
  \end{center}
  \caption{Parallel speed up of new algorithm and its components}
  \label{fig:parallelscaling}
\end{figure}

\figref{fig:parallelscaling} shows that the speed-up of the creation of
eigenvalues with processor count is super-linear.  This is due to the fact
that with increasing processor count the number of eigenvectors which can be
held in cache on each processor increases as the local sub-vectors become
smaller.  The net result is a super-linear speed-up of the entire algorithm
with processor count.

In order to illustrate the efficiency of our implementation of our new variant
of the LANSO algorithm to find low-lying eigenmodes of the fermion matrix we
compare it with the current state--of--the--art, the Chroma implementation of
the Kalkreuter--Simma algorithm described in ~\cite{Kalkreuter:1995mm}.  This
method uses a conjugate gradient (CG) method to minimize the Ritz functional
\(\mu(\vect z) = (\vect z, \ma A\vect z)/\|\vect z\|^2\) with \(\ma A =
(\gamma_5 \ma D)^2\), where \(\vect z\) is deflated with respect to all
previously computed eigenvectors.  The CG minimization alternates with a
diagonalization of \(\gamma_5 \ma D\) on the subspace of computed eigenvectors
to separate eigenvalues of \(\gamma_5 \ma D\) of different sign but the same
magnitude, taking into account that we may not know the full degenerate
subspaces.

Comparing like-with-like for the various methods of determining eigenpairs is
not completely straightforward as one has to consider some kind of tolerance
within which the eigenvalues are determined.  In the case of the
Kalkreuter--Simma algorithm convergence is specified by stopping criteria on
the CG method, whereas in our new algorithm we determine whether a Ritz pair
\((\theta,\vect y)\) has converged by looking at the bottom component of the
Ritz vector.  Moreover, we continue to refine the eigenpairs at each pause, so
their accuracy improves: we could deflate with respect to sufficiently good
eigenvectors but we have not studied this option.

\begin{figure}[!ht]
  \begin{center}
    \epspdffile{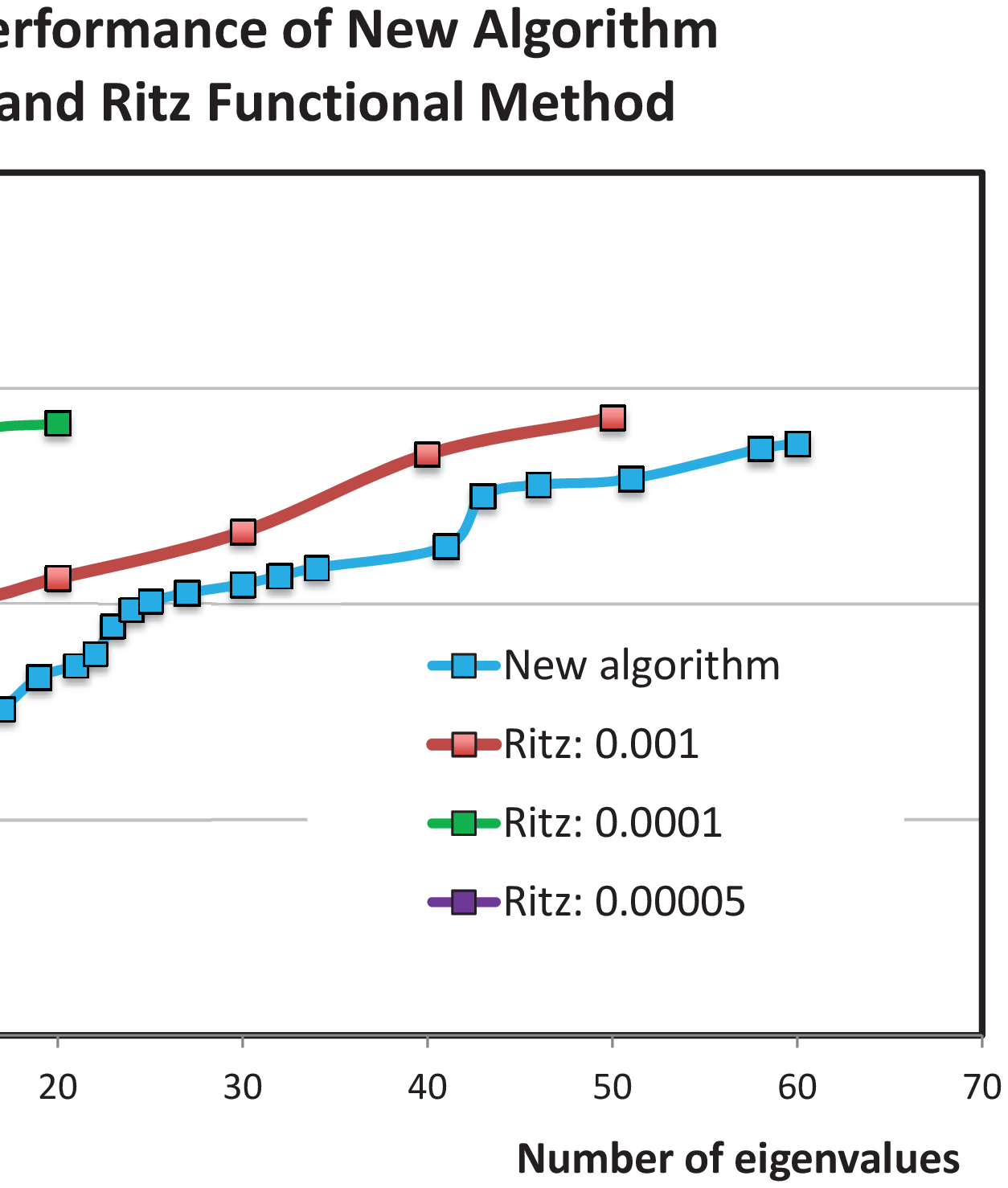}
  \end{center}
  \caption{Comparison of our algorithm and the Ritz functional
    method~\cite{Kalkreuter:1995mm,Bunk:1996kt} implemented in Chroma and run
    on HECToR as a function of the number of small magnitude eigenpairs of
    \(\gamma_5 \ma D\) found.  Results are shown for different residual values
    for the Ritz method; the corresponding errors for our method are always
    smaller than the best Ritz functional estimates, and decrease as the
    Krylov space grows.}
  \label{fig:ritz}
\end{figure}

We compare the results studying the norm of the residual vector \(\|(\ma A -
\theta)\vect y\|\).  We adjust the relevant stopping criteria and tolerances
until we see similar magnitudes of this norm and then compare the result in
terms of the overall computation time: the results are in \figref{fig:ritz}.

\section{Conclusions}

We have introduced a new algorithm to determine the eigenpairs of large
Hermitian matrices based on the LANSO method of Parlett and Scott, and
implemented and tested it on a realistic large-scale computation in lattice
QCD.  Our algorithm differs in two ways from LANSO: it only determines
eigenpairs within a specified region of the spectrum, as this is all that is
needed, and it uses a new \(\sigma\) bound to trigger ``pauses'' at which Ritz
pairs are computed and selective reorthogonalization performed.  We found that
this reduces the number of such pauses significantly, and moreover far less
work is required as we only need to construct the eigenvectors we are
interested in.  Our method compares very favourably with the methods that are
currently in use, and promises to be useful for other problems such as
``low-mode averaging'' in QCD calculations as well as in applications in other
areas.  We have indicated several possible improvements that could be studied
in future.

\section{Acknowledgements}

% QCD support grant

We would like to thank B\'alint Jo\'o for his help with Chroma.  We gratefully
acknowledge the support of the Centre for Numerical Algorithms and Intelligent
Software (EPSRC EP/G036136/1) together with STFC (ST/G000522/1) in the
preparation of this work.

This work made use of the facilities of HECToR, the UK's national
high-performance computing service, which is provided by UoE HPCx Ltd at the
University of Edinburgh, Cray Inc and NAG Ltd, and funded by the Office of
Science and Technology through EPSRC's High End Computing Programme.

\bibliographystyle{elsart-num}
\bibliography{lanso}
\end{document}